\documentclass[aps,prc,
reprint,
twocolumn,
showpacs,
preprintnumbers,
floatfix,
superscriptaddress]{revtex4-1}

\usepackage{amsmath}
\usepackage{graphicx}
\usepackage{rotating} % Rotating table

\newcommand{\tra}[4]{\ensuremath{{#1}_{#2}^{+} \rightarrow {#3}_{#4}^{+} }}

\usepackage{lineno}

\begin{document}

\title{$E0$ transition strength in stable Ni isotopes}


\author{L.J.~Evitts}
\altaffiliation{Present address: Nuclear Futures Institute, Bangor University, Bangor, Gwynedd, LL57 2DG, UK}
\affiliation{TRIUMF, 4004 Wesbrook Mall, Vancouver, B.C., V6T 2A3, Canada}
\affiliation{Department of Physics, University of Surrey, Guildford, Surrey, GU2 7XH, United Kingdom}

\author{A.B.~Garnsworthy}
\email[]{garns@triumf.ca}
\affiliation{TRIUMF, 4004 Wesbrook Mall, Vancouver, B.C., V6T 2A3, Canada}

\author{T.~Kib\'{e}di}
\affiliation{Department of Nuclear Physics, Research School of Physics and Engineering, The Australian National University, Canberra, ACT 2601, Australia}
\author{J.~Smallcombe}
\altaffiliation{Present address: Oliver Lodge Laboratory, The University of Liverpool, Liverpool, L69 7ZE, UK}
\affiliation{TRIUMF, 4004 Wesbrook Mall, Vancouver, B.C., V6T 2A3, Canada}
\author{M.W.~Reed}
\altaffiliation{Present address: Mettler-Toledo Safeline X-Ray Ltd, Royston, Herts, SG8 5HN, United Kingdom}
\author{A.E.~Stuchbery}
\author{G.J.~Lane}
\author{T.K.~Eriksen}
\altaffiliation{Present address: Department of Physics, University of Oslo, P. O. Box 1048 Blindern, N-0316 Oslo, Norway}

\author{A.~Akber}
\affiliation{Department of Nuclear Physics, Research School of Physics and Engineering, The Australian National University, Canberra, ACT 2601, Australia}
\author{B.~Alshahrani}
\affiliation{Department of Nuclear Physics, Research School of Physics and Engineering, The Australian National University, Canberra, ACT 2601, Australia}
\affiliation{Department of Physics, King Khalid University, Abha, Kingdom of Saudi Arabia}
%\author{B.A.~Brown}
%\affiliation{National Superconducting Cyclotron Laboratory, Michigan State University, East Lansing, Michigan 48824, USA}
%\affiliation{Department of Physics and Astronomy, Michigan State University, East Lansing, Michigan 48824, USA}
\author{M.~de Vries}
\author{M.S.M.~Gerathy}
\affiliation{Department of Nuclear Physics, Research School of Physics and Engineering, The Australian National University, Canberra, ACT 2601, Australia}
\author{J.D.~Holt}
\affiliation{TRIUMF, 4004 Wesbrook Mall, Vancouver, B.C., V6T 2A3, Canada}
\author{B.Q.~Lee}
\altaffiliation{Present address: Department of Physics, University of Oxford, Oxford, OX1 3PJ, United Kingdom}
\author{B.P.~McCormick}
\author{A.J.~Mitchell}
\affiliation{Department of Nuclear Physics, Research School of Physics and Engineering, The Australian National University, Canberra, ACT 2601, Australia}
\author{M.~Moukaddam}
\altaffiliation{Present address: PHC-DRS/Universit\'{e} de Strasbourg, IN2P3-CNRS, UMR 7178, F-67037, Strasbourg, France}
\affiliation{TRIUMF, 4004 Wesbrook Mall, Vancouver, B.C., V6T 2A3, Canada}
\author{S.~Mukhopadhyay}
\affiliation{Departments of Chemistry and Physics \& Astronomy, University of Kentucky, Lexington, Kentucky, 40506-0055, USA}
\author{N.~Palalani}
\altaffiliation{Present address: Department of Physics, University of Botswana, 4775 Notwane Rd, Gaborone, Botswana}
\author{T.~Palazzo}
\affiliation{Department of Nuclear Physics, Research School of Physics and Engineering, The Australian National University, Canberra, ACT 2601, Australia}
\author{E.E.~Peters}
\author{A.P.D.~Ramirez}
\affiliation{Departments of Chemistry and Physics \& Astronomy, University of Kentucky, Lexington, Kentucky 40506-0055, USA}
%\author{S.R.~Stroberg}
%\altaffiliation{Present address: Physics Department, Reed College, 3203 SE Woodstock Blvd., Portland, OR 97202-8199, USA}
%\affiliation{TRIUMF, 4004 Wesbrook Mall, Vancouver, B.C., V6T 2A3, Canada}
\author{T.~Tornyi}
\affiliation{Department of Nuclear Physics, Research School of Physics and Engineering, The Australian National University, Canberra, ACT 2601, Australia}
\author{S.W.~Yates}
\affiliation{Departments of Chemistry and Physics \& Astronomy, University of Kentucky, Lexington, Kentucky, 40506-0055, USA}

\date{\today}

\begin{abstract}
%JTS
Excited states in $^{58,60,62}$Ni were populated via inelastic proton scattering at the Australian National University as well as via inelastic neutron scattering at the University of Kentucky Accelerator Laboratory. The Super-e electron spectrometer and the CAESAR Compton-suppressed HPGe array were used in complementary experiments to measure conversion coefficients and $\delta(E2/M1)$ mixing ratios, respectively, for a number of $2^+ \rightarrow 2^+$ transitions. The data obtained were combined with lifetimes and branching ratios to determine $E0$, $M1$, and $E2$ transition strengths between $2^+$ states. The $E0$ transition strengths between $0^+$ states were measured using internal conversion electron spectroscopy and compare well to previous results from internal pair formation spectroscopy. The $E0$ transition strengths between the lowest-lying $2^+$ states were found to be consistently large for the isotopes studied.
\end{abstract}


\maketitle

%\linenumbers

\section{Introduction \label{sect:intro}}

The strength of an electric monopole ($E0$) transition, $\rho^2 (E0)$, can be directly related to the difference in deformation between the initial and final states, as well as the degree of mixing between them. Evidence of significant $E0$ strength has been associated with shape coexistence \cite{Heyde.83.1467}.  The presence of an $E0$ transition can also be used as a test of various nuclear models, such as the axially symmetric quadrupole rotor or the spherical vibrator model, in which selection rules are placed on $E0$ transitions \cite{Wood1999}. 

Single $\gamma$-ray emission is forbidden for an $E0$ transition as a photon must carry away at least 1$\hbar$ of angular momentum. 
While $E2$ transition matrix elements can be extracted in Coulomb excitation studies, the $E0$ component is not directly accessible in this approach.
Therefore, there is a need to employ electron spectroscopy for the determination of $E0$ transition strengths. 
%These measurements are susceptible to many additional sources of background not present in $\gamma$-ray spectroscopy. 

The number of $E0$ transition strengths known experimentally is quite low in comparison to measurements of $E2$ transitions, as a result of a number of experimental challenges.  
Comparing the experimental data available from the three most recent compilations, one finds that there are 447, 87 and 14 evaluated values reported for B($E2$ : \tra{2}{1}{0}{1}) \cite{Pritychenko20161}, $\rho^2 (E0$ : \tra{0}{2}{0}{1}) \cite{Kibedi2005} and $\rho^2 (E0$ : \tra{2}{2}{2}{1}) \cite{Wood1999} transition strengths, respectively.  
These statistics are expected to change as there have been a number of advances and a rejuvenation of the detection systems being employed for electron and positron spectroscopy worldwide in recent years \cite{Battaglia2016,Cox2017,Papadakis2015,Pakarinen2014,Ketelhut2014,Perkowski2014}. One area where data are still particularly lacking is a characterization of $E0$ transition strengths between states of $J>0$ in spherical nuclei. This deficiency is the motivation for the present study of the nickel isotopes \cite{Leoni.118.162502, Suchyta.89.021301, Prokop.92.061302}. Detailed muonic X-ray measurements \cite{Shera1976} and optical spectroscopy \cite{Steudel1980} indicate that the ground states of these isotopes are spherical with little variation.

Previous experimental work has yielded the $\rho^2 (E0)$ values between 0$^+$ states in $^{58, 60, 62}$Ni \cite{Passoja1981, Warburton197138}.  Two previous experiments were performed with the ($p$, $p^\prime$) reaction and $E0$ transition strengths were determined by observing the electron-positron pairs emitted in internal pair formation ($\pi$) decay.  There has been no previous work in determining $\rho^2 (E0)$ values between $J^\pi=0^+$ states in these nuclides through the measurement of conversion electrons.  

There is a notable deficiency of $\rho^2 (E0)$ values measured between $J_i^{\pi} = J_f^{\pi} \neq 0^+$ states across the entire chart of nuclides and especially in light- and medium-mass nuclei; none have been previously measured in the Ni isotopes.  
As the $E0$ strength is closely related to the change in shape of a nucleus, there is a need for values to be measured in a wide range of nuclei.  Determining the $E0$ strength between $J_i^{\pi} = J_f^{\pi} \neq 0^+$ states requires the experimental determination of a number of quantities, often necessitating different experimental setups. The experimental quantities include the $E2$/$M1$ mixing ratio, the parent state half-life, the internal conversion coefficient, and the transition branching ratio.

In this article, we report details and results from measurements of $E0$ transition strengths between $2^+$ states in $^{58,60,62}$Ni. Initial results from this experimental study, focusing on only the \tra{2}{2}{2}{1} transitions, were published in Ref. \cite{Evitts_PLB}. The measurements were performed at the Australian National University (ANU) and the University of Kentucky Accelerator Laboratory (UKAL).

%\section{Experiments and Analysis at ANU \label{sect:ExpDetailsANU}}
\section{Experiments and Analysis following (p, p') reactions \label{sect:ExpDetailsANU}}
Two experiments were carried out at the Heavy Ion Accelerator Facility at the ANU.  Proton beams between 4.7 and 9.2\,MeV were provided by the 14UD pelletron.  Self-supporting targets with a thickness of 1.4\,mg/cm$^2$ for $^{58}$Ni and 1.3\,mg/cm$^2$ thickness for $^{60, 62}$Ni were used. The isotopic enrichments for the $^{58,60,62}$Ni foils were 99.1\%, 99.8\% and 98.8\%, respectively. The same set of targets was used in all measurements.

\subsection{Apparatus \label{sect:Apparatus}}
The CAESAR array, composed of nine Compton-suppressed HPGe detectors, was used for measurements of angular distributions of $\gamma$ rays. 
Data were collected for approximately 2 hours with each target at a beam intensity of 5-10\,nA.

\begin{figure}[!ht]
	\centering
	\includegraphics[angle=-90,width=0.9\linewidth]{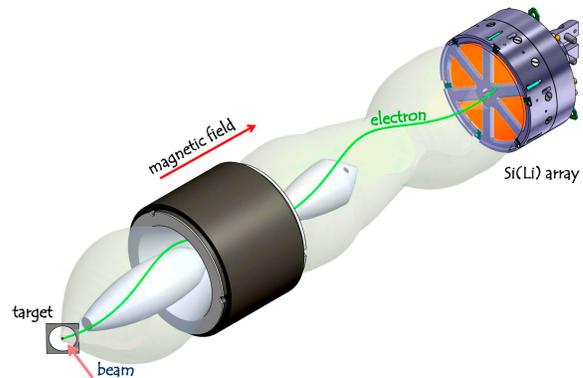}
	\caption{Schematic diagram (not to scale) of the superconducting electron (Super-e) spectrometer at the ANU. The spectrometer was developed for electron-positron pair spectroscopy, but here was used to collect electron singles events.}
	\label{fig:supere_schematic}
\end{figure}

The second experimental setup was the superconducting electron spectrometer, Super-e \cite{Kibedi1990}, which is composed of a solenoid magnet and thick lithium-drifted silicon [Si(Li)] detector. The configuration of the Super-e is shown in Fig. \ref{fig:supere_schematic}. A Compton-suppressed HPGe detector was placed close to the target to allow for simultaneous measurements of $\gamma$ rays.  The proton beam was incident on the self-supporting target tilted at 45$^{\circ}$ to the beam. Unreacted beam continues on to a Faraday cup in the beam dump for the purpose of monitoring the beam current. The proton beam was provided at up to 800\,nA for approximately 6-12 hours on each target.

\begin{figure}[ht]
	\centering
	\includegraphics[width=0.93\linewidth, keepaspectratio, angle=90, trim={0 0 0 0cm}, clip]{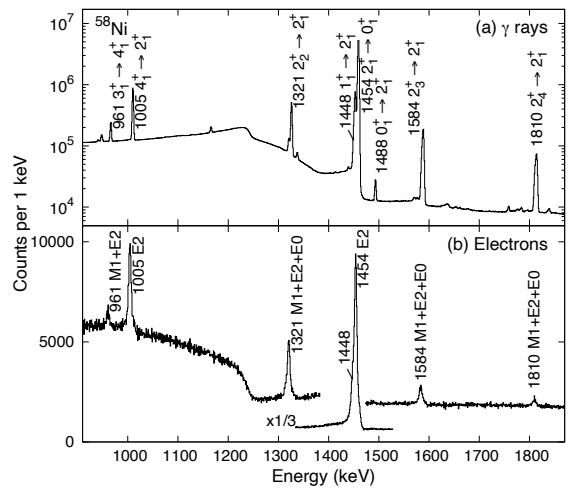}
	\caption{Gamma-ray (a) and electron (b) energy spectra collected for the $^{58}$Ni targets.
    }
	\label{fig:spectra58}
\end{figure}

\begin{figure}[ht]
	\centering
	\includegraphics[width=0.93\linewidth, keepaspectratio, angle=90, trim={0 0 0 0cm}, clip]{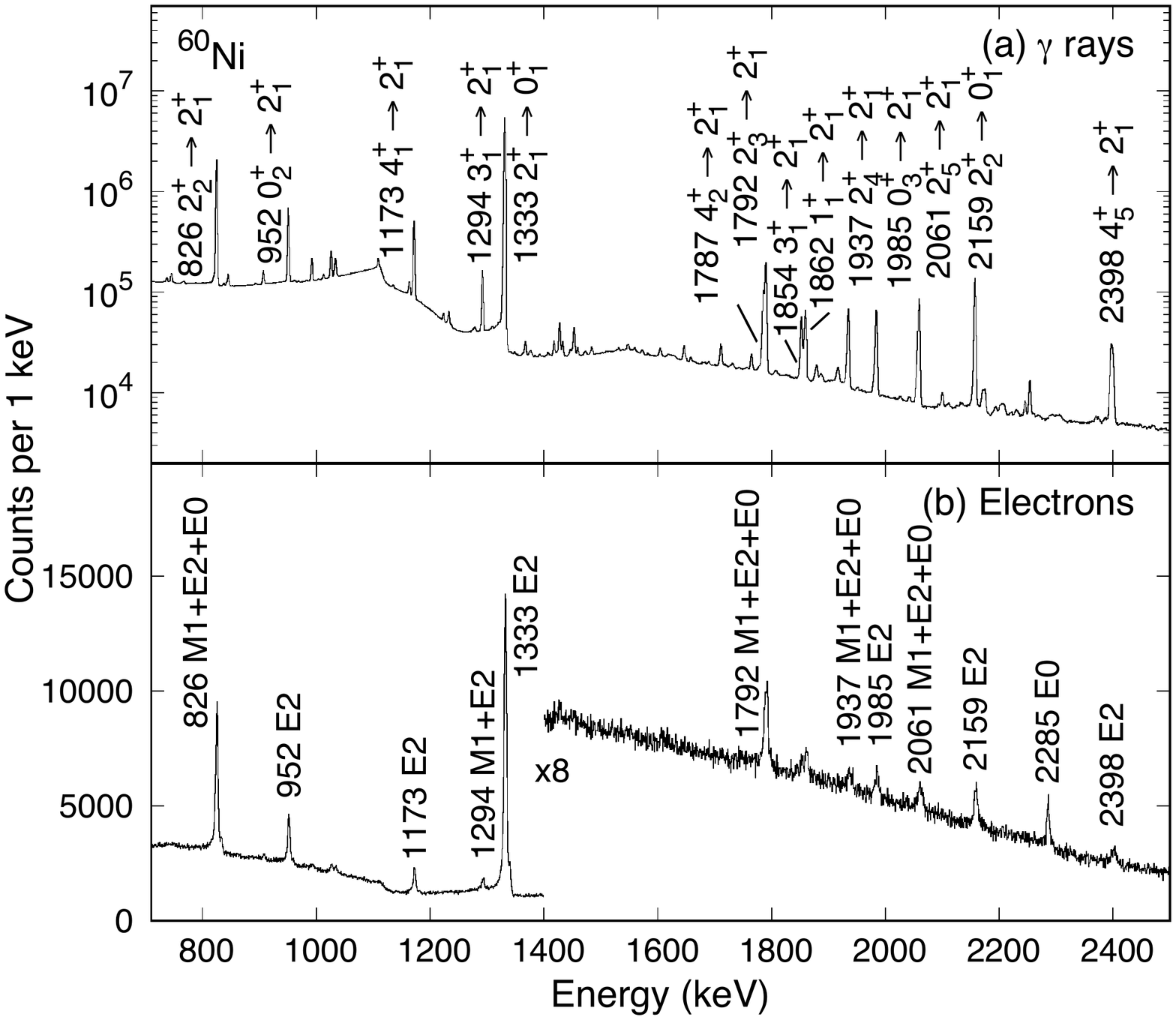}
	\caption{Gamma-ray (a) and electron (b) energy spectra collected for the $^{60}$Ni targets.
    }
	\label{fig:spectra60}
\end{figure}

\begin{figure}[ht]
	\centering
	\includegraphics[width=0.93\linewidth, keepaspectratio, trim={0 0 0 0cm}, clip]{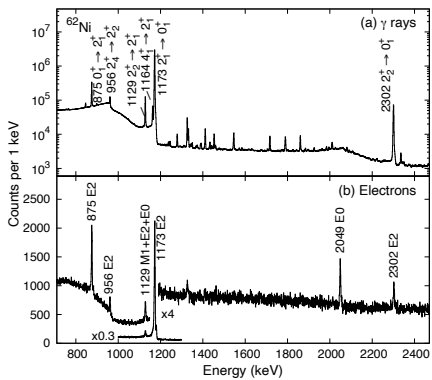}
	\caption{Gamma-ray (a) and electron (b) energy spectra collected for the $^{62}$Ni targets.
    }
	\label{fig:spectra62}
\end{figure}

Electrons emitted from the target are transported by the magnetic field of the superconducting solenoid magnet around two baffles and through a diaphragm in order to be incident on a set of six 9\,mm thick Si(Li) detectors located 35\,cm from the target. The geometry is such that each electron of a given energy ($E$) must complete 2.5 helical orbits in the magnetic field before reaching the detector. During an experiment, the magnetic field was swept over a range between the minimum and maximum set values. The period of time spent at each step of the magnetic field setting in the cycle was variable so that the integrated charge of the proton beam recorded in the Faraday cup was the same for each field value. The peak-to-total ratio in the electron energy spectrum was improved by gating on the magnetic field value that is recorded in the data stream. 
As the energy of the transported electron is related to the momentum window defined by the magnetic field, the selection of only events in this window can reduce the contribution of background and of events in which the full electron energy has not been recorded in the Si(Li) detector. Gamma-ray and electron energy spectra collected from the Super-e detector are shown in Figs. \ref{fig:spectra58}, \ref{fig:spectra60} and \ref{fig:spectra62} for each of the $^{58,60,62}$Ni isotopes.

\begin{figure}[!ht]
	\centering
	\includegraphics[width=0.9\linewidth]{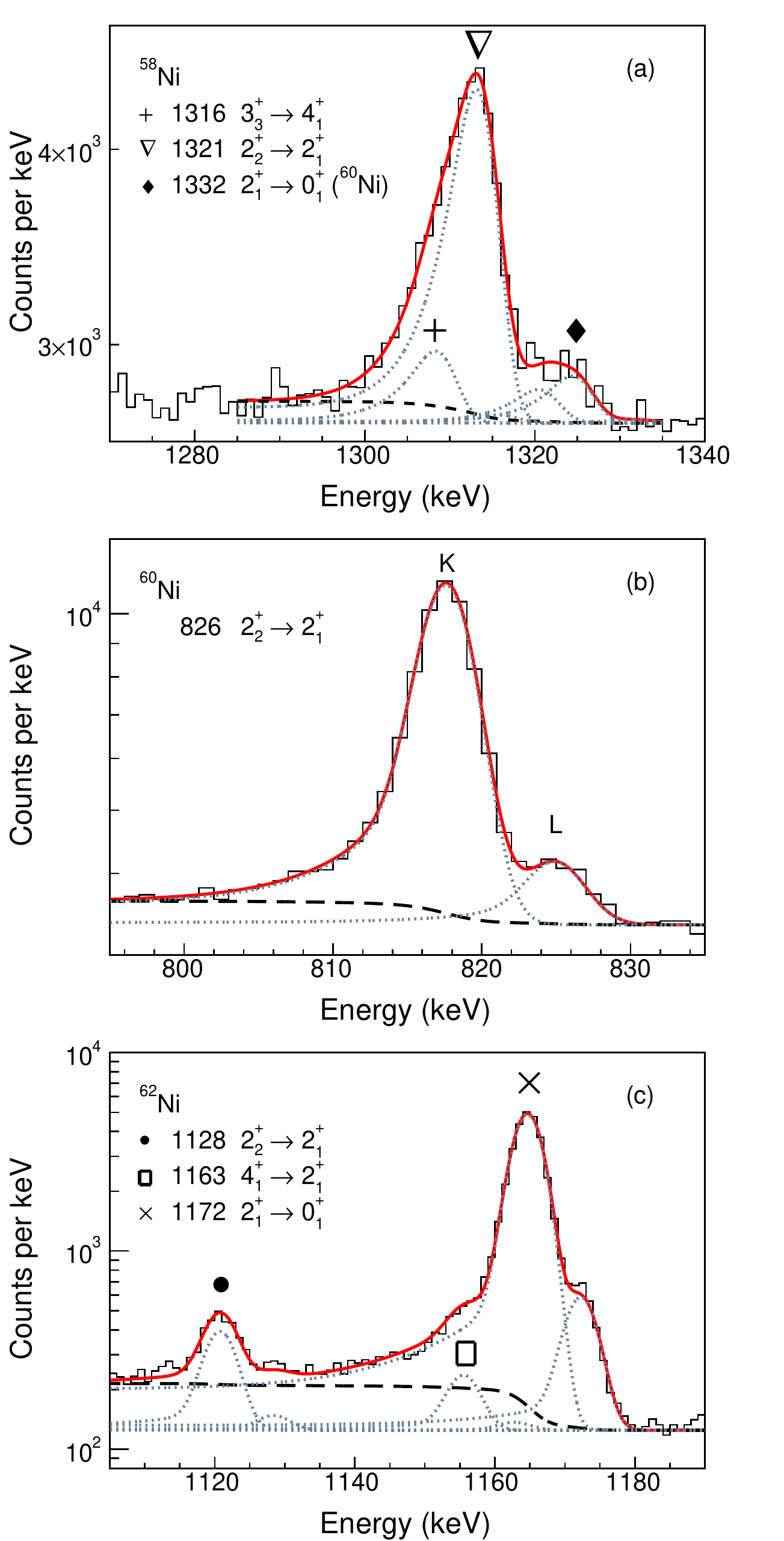}
	\caption{Peak fitting of the \tra{2}{2}{2}{1} transitions in the electron spectra collected with Super-e for the (a) $^{58}$Ni, (b) $^{60}$Ni and (c) $^{62}$Ni target. The background fit is shown by a black dashed line, each individual peak is shown by a grey dotted line and the total fit is shown by a full red line.  For each transition, there are two peaks corresponding to the K and L electrons.  Each fit has a reduced $\chi^2$ value of (a) 1.1, (b) 1.0 and (c) 1.2.
    }
	\label{fig:spectra58-60-62}
\end{figure}

The $\gamma$ rays emitted from the target were detected by a single Compton-suppressed HPGe detector located outside the chamber, approximately 50\,cm from the target. The $\gamma$-ray energy spectrum was used for normalization of the electron data and in the measurement of internal conversion coefficients.

\subsection{Calibration source preparation \label{sect:Source}}
The radionuclide $^{170}$Lu decays by electron capture with a half-life of 2\,days to excited states in $^{170}$Yb and subsequently emits a large number of $\gamma$ rays and conversion electrons between 20\,keV and 3.4\,MeV. This large number of discrete transitions in this decay make $^{170}$Lu an excellent calibration source for the determination of the relative efficiency of both $\gamma$-ray and electron detectors. 

In order to produce a $^{170}$Lu source, a $^{171}$Yb foil of 95.1\,\% isotopic enrichment and a thickness of 2\,mg/cm$^2$ was irradiated in a shielded location at the ANU.  Over a period of 16 hours, an 18\,MeV proton beam with a current of 25\,nA impinged upon the target.  The beam current was limited by the levels of radiation permitted in the experimental hall.

The internal conversion coefficients of the majority of transitions emitted following the decay of $^{170}$Lu have been measured with good accuracy \cite{Camp1972,NucData170}. The use of this calibration source is also discussed in Ref. \cite{Eriksen2017}. This $^{170}$Lu source is particularly useful in the case of electron detectors as there are few long-lived radionuclides suitable as discrete-energy electron calibration sources, especially at higher electron energies.

\subsection{Efficiency calibrations \label{sect:EffAnalysis}}
The relative efficiencies of the HPGe detectors in the CAESAR array were calibrated over the energy region of interest using $^{56}$Co and $^{170}$Lu sources. 

The theoretical transport efficiency of the Super-e spectrometer is calculated as,

\begin{equation}
\label{eq:trans_eff}
y(E) = \frac{A}{m_ec^2} \cdot \sqrt{(E^2 + 2m_ec^2E)},
\end{equation}

\noindent where $A$ is a normalizing factor, $m_e$ is the electron rest mass, $c$ is the speed of light and $E$ is the kinetic energy of an electron in keV. The normalizing factor can take on three values corresponding to the lower and upper limits, and optimum transmission for a given energy.

At higher energies, consideration of the detector response must also be taken into account, in addition to the transport efficiency. A GEANT4 \cite{Allison2016186} simulation was used to determine the ratio of events that deposit their full energy in the detector to the total number of electrons that are incident on the detector. 
The inputs to this simulation were the electron momentum vectors resulting from a simulation of the trajectories through the spectrometer in order to correctly consider the variation in incident angle of the electrons reaching the detector surface.
The detector response determined from the simulation is combined with the transport efficiency of Eq. \eqref{eq:trans_eff} to obtain the total efficiency. The total efficiency was normalized to the data from the $^{170}$Lu source. The energy dependence of the detector efficiency is only significant above 2\,MeV, thus for all transitions studied in this work, the total efficiency is equal to the transport efficiency.

\subsection{Angular distributions \label{sect:AngDistAnalysis}}
The angular distributions of $\gamma$ rays can be used to determine the $E2/M1$ mixing ratio, $\delta$, for transitions of mixed multipolarity by fitting the function

\begin{equation}
\label{eq:ang_dist}
W(\theta) = N \cdot [1 + \alpha_2 Q_2 A_2 P_2(\text{cos } \theta) + \alpha_4 Q_4 A_4 P_4(\text{cos } \theta) ],
\end{equation}

\noindent where $N$ is a normalization parameter, $Q_k$ are finite solid angle correction factors, $P_k(x)$ are the Legendre polynomials of the k$^{th}$ order,  $\alpha_k$ are the attenuation coefficients, which depend on the degree of alignment of the parent state, and $A_k$ are the angular distribution coefficients, which depend on the parent spin and the mixing ratio of the transition \cite{Yamazaki67}.  

There can be variations in the physical position of the beam incident on each of the targets as well as with the positioning of the radioactivity in the calibration source. Such differences modify the apparent angle of each detector and the emitted radiation. Following the efficiency calibration, the apparent angle of each detector was determined separately for each target by a chi-squared minimization using the angular distribution of known pure $E2$ transitions emitted from the target nuclei. 
Deviations were at most a few degrees from the nominal angles determined from physical measurements of detectors with respect to the beam axis.

The parameter $Q_k$ is a solid-angle correction factor for the finite size of the HPGe detectors that depends on the size, orientation and opening angle of the crystal exposed by the collimator \cite{Krane71}. The geometrical solid angle attenuation coefficients for CAESAR have been previously evaluated to be $Q_{2}$=0.98 and $Q_{4}$=0.94 \cite{Andressen_PhD}. The uncertainty in the $Q_k$ coefficients does not exceed 1\%, which more than covers their dependence on $\gamma$-ray energy.

The alignment of the parent state for each transition of interest was determined by fitting the angular distribution of the competing $\gamma$ ray from the parent state to the $0^+$ ground state with the function of Eq. \eqref{eq:ang_dist}.  As this is a pure $E2$ transition, the alignment coefficients, $\alpha_k$, are determined by fixing the other angular distribution coefficients, $A_k$, to the theoretical values.  The alignment coefficients were then adopted in determining the mixing ratio of the mixed transitions.  The values of $\delta$ are taken from the minima in a plot of $\chi^2$ versus $\delta$ and the 1$\sigma$ limits are defined by the range of $\chi^2$+1 \cite{Avni76,Robinson1990}.

\subsection{Internal conversion coefficients and $\rho^2$($E0$) values \label{sect:ICCAnalysis}}
Accurate peak fitting is essential in the determination of yields for transitions that lie close in energy and are, therefore, overlapping in the electron spectrum. The shape parameters of the electron peaks, which in this case depend primarily on the energy of the electron and detector effects, were fixed by fitting transitions of similar energy in an $^{54}$Fe dataset that was collected during the same beam time. The contribution to peak shape from energy straggling in the target or energy broadening from in-flight emission is minimal in this study and was not specifically considered in the fitting of electron peaks. In the case of pure $E2$ transitions, it was possible to also fix the expected ratio of conversion from the K and L atomic subshells. The change in efficiency between the K and L energies ($\sim8$\,keV) is negligible. 

The electric monopole transition strength, $\rho^2 (E0)$, can be determined from \cite{Kibedi2005}

\begin{equation}
\rho^2 (E0) = \frac{1}{\Omega_K (E0) \cdot \tau_K (E0)},
\end{equation}

\noindent where $\Omega_K (E0)$ is the electronic factor obtained from atomic theory \cite{Kibedi2008} and $\tau_K (E0)$ is the partial mean lifetime of the $E0$ component converted in the K shell.  The $\tau_K (E0)$ is calculated using the relative branching ratio of the $E0$ transition, $\lambda_{E0}$, to the sum of all available decay modes, $\sum_i \lambda_i$, from the parent state, i.e.,

\begin{equation}
\label{eq:tau_from_lamb}
\tau_k (E0) = \frac{\sum_i \lambda_i}{\lambda_{E0}} \cdot \frac{T_{1/2}}{\textnormal{ln} (2)},
\end{equation}

\noindent where $T_{1/2}$ is the half-life of the parent state. Each contribution, such as the mixing ratio, if not measured in the present experiment, can be calculated from experimental data available in the literature. 
A number of the input values, particularly the parent half-life and mixing ratios, have asymmetric uncertainties.  
These asymmetric values lead to an overestimated uncertainty in the final value when calculated through standard error propagation.  
As such, the final value and uncertainties in this work were determined through a Monte Carlo method from which the median value and the 1\,sigma (68\%) confidence interval are presented.

%\section{Experiments and Analysis following (n, n') reactions \label{sect:ExpDetailsUKAL}}
\section{Experiments and Analysis at the UKAL \label{sect:ExpDetailsUKAL}}

Inelastic neutron scattering (INS) measurements were performed at the University of Kentucky Accelerator Laboratory (UKAL), which houses a 7\,MV Van de Graaff accelerator capable of producing high-quality pulsed and bunched beams. Nearly monoenergetic neutrons were produced via the $^3$H(p,n)$^3$He reaction using a gas cell containing approximately an atmosphere of tritium gas.  A single $\approx$50\% efficient HPGe detector surrounded by an annular bismuth germanate (BGO) shield for Compton suppression was used for $\gamma$-ray detection.  Time-of-flight gating was also employed to reduce the background for the prompt spectra.  For the measurements, a cylindrical scattering sample of Ni metal of natural abundance, 45.94\,g mass, 1.84\,cm height, and 1.88\,cm diameter was used.

\begin{figure}[!ht]
\centering
\includegraphics[width=1.0\linewidth]{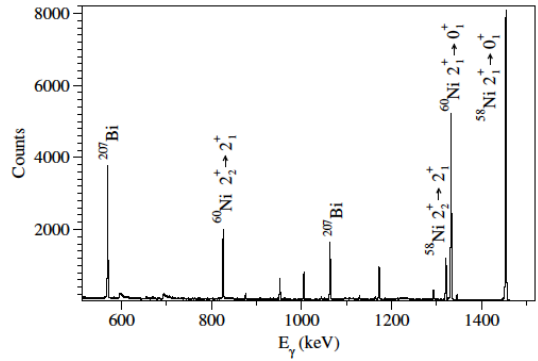}
\caption{\label{fig:UKAL-spec}Gamma-ray energy spectrum of inelastic neutron scattering on a natural Ni target. A $^{207}$Bi source was placed near the HPGe detector during the measurements to provide an ``online" energy calibration.}
\end{figure}

\begin{figure}[!ht]
\centering
\includegraphics[width=1.0\linewidth, keepaspectratio, trim={0 0 0 0cm}, clip]{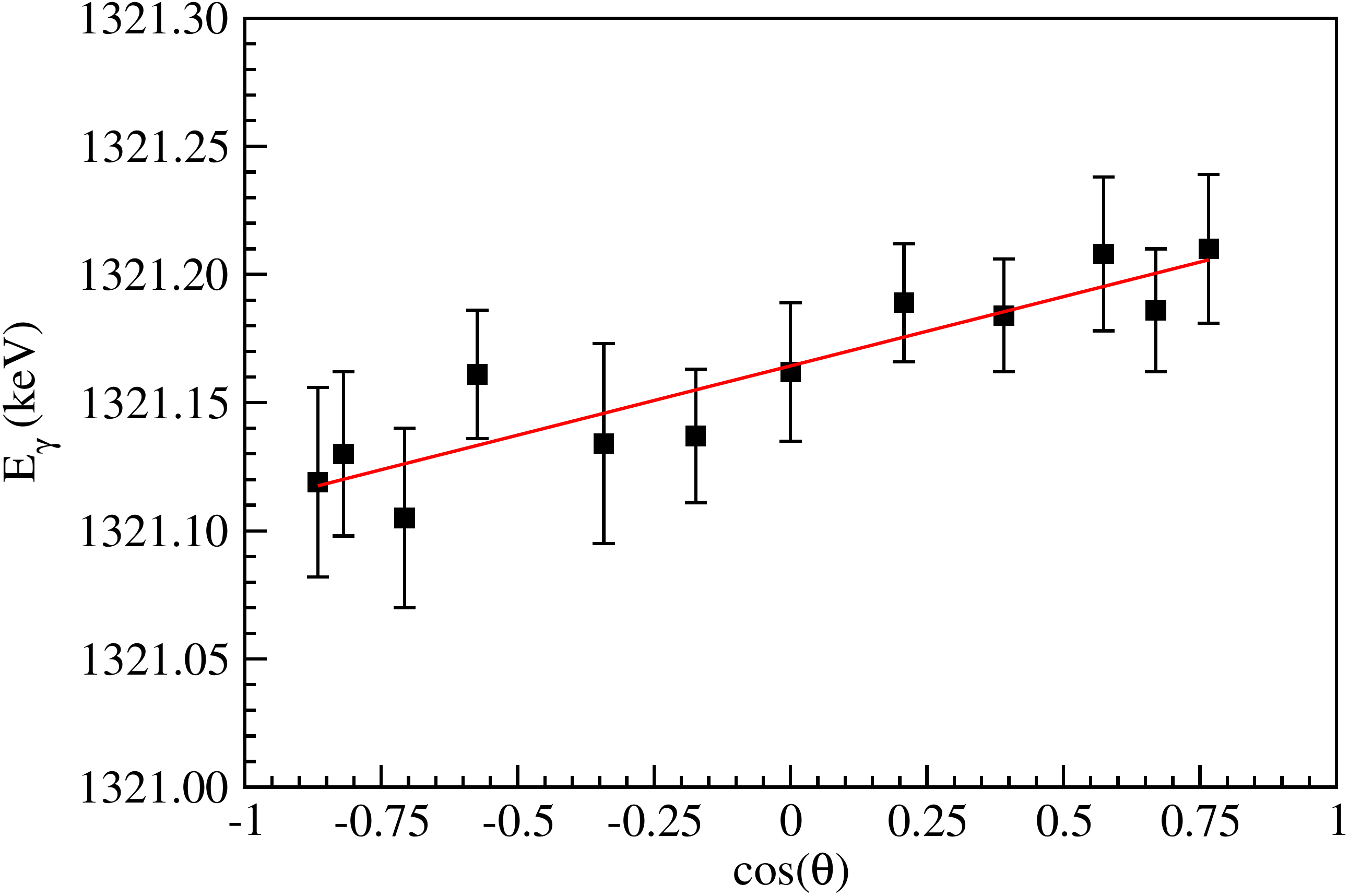}
\caption{Doppler-shift data for the 1321\,keV $\gamma$ ray from the 2775\,keV 2$^+_2$ level in $^{58}$Ni. The line is a linear fit to the data. }~\label{fig:tau}
\end{figure}

\begin{figure}[!ht]
\centering
\includegraphics[width=1.0\linewidth, keepaspectratio, trim={0 0 0 0cm}, clip]{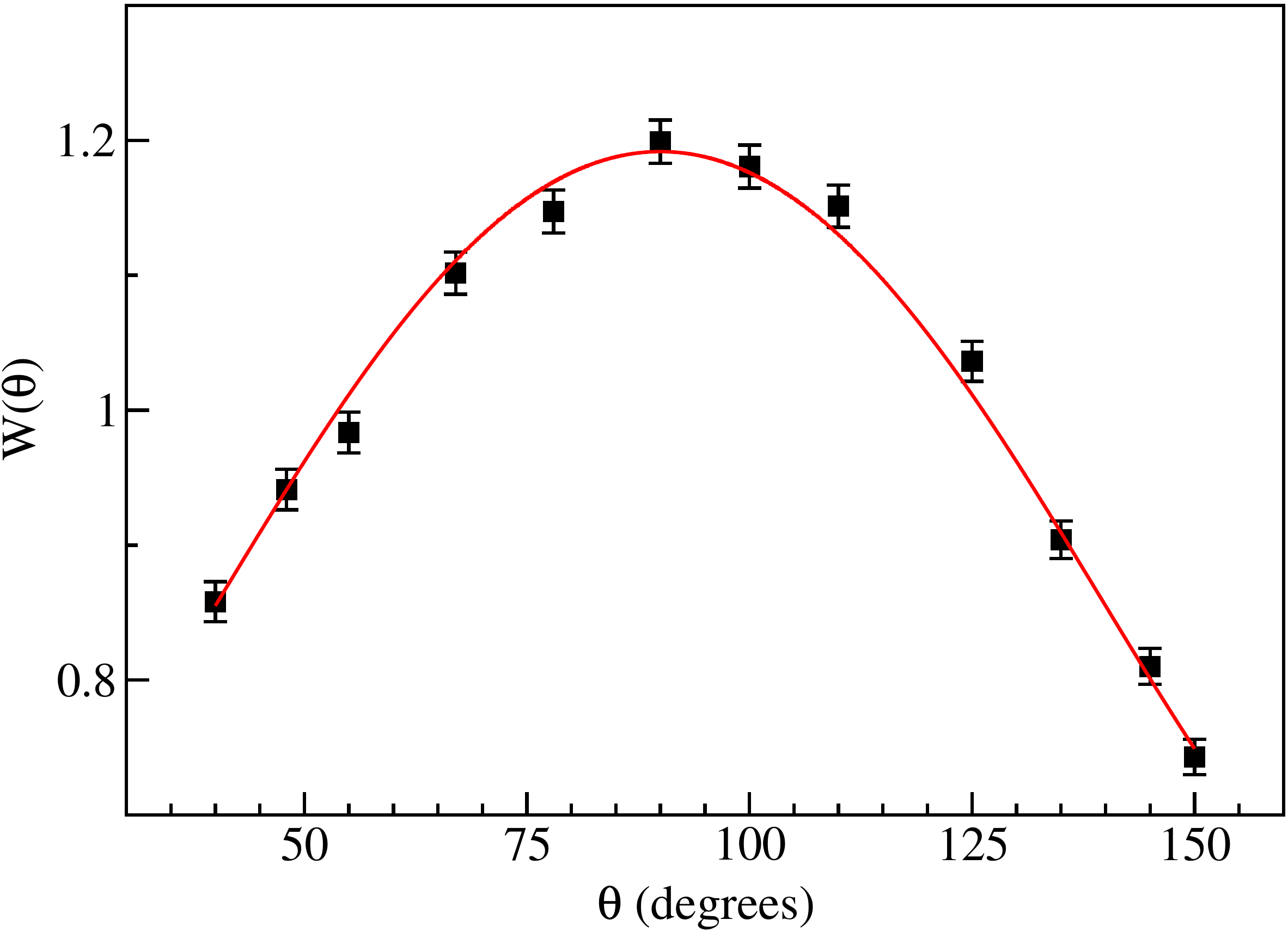}
\caption{Gamma-ray angular distribution of the 1321\,keV $\gamma$ ray from the 2775\,keV 2$^+_2$ level in $^{58}$Ni. The line is a Legendre polynomial fit to the data. }~\label{fig:AD}
\end{figure}

Angular distribution measurements were performed for incident neutron energies of 2.42 and 2.90\,MeV.  The detector was rotated between 40 and 150$^\circ$ with respect to the incident beam direction.  
A $^{207}$Bi radioactive source was placed near the HPGe detector during the INS measurements, providing an ``online" internal energy calibration, while $^{226}$Ra was used offline for non-linearity and efficiency corrections.  
From these data, level lifetimes were extracted using the Doppler-shift attenuation method (DSAM) \cite{Bel96}.  An example of the Doppler-shift data is shown in Fig. \ref{fig:tau}.  From the slope of the linear fit to the data, the experimental attenuation factor, $F(\tau)$, was extracted and compared with calculations using the Winterbon formalism \cite{winterbon} in order to determine the lifetime.  The multipole mixing ratio ($\delta$) was extracted by comparing the fitted Legendre polynomial coefficients ($a_2$ and $a_4$) for the angular distribution to those calculated by the statistical model code CINDY \cite{cindy} as a function of $\delta$.  An example of a $\gamma$-ray angular distribution is shown in Fig. \ref{fig:AD}.  Complete details of the analysis methods are described in a previous study of $^{62}$Ni at the UKAL \cite{chakraborty2011}.

\section{Results and Discussion\label{sect:results}}
\subsection{$E2$/$M1$ mixing ratios from angular distributions of $\gamma$ rays \label{sect:MR}}

\begin{figure}[!ht]
\centering
\includegraphics[width=1.0\linewidth]{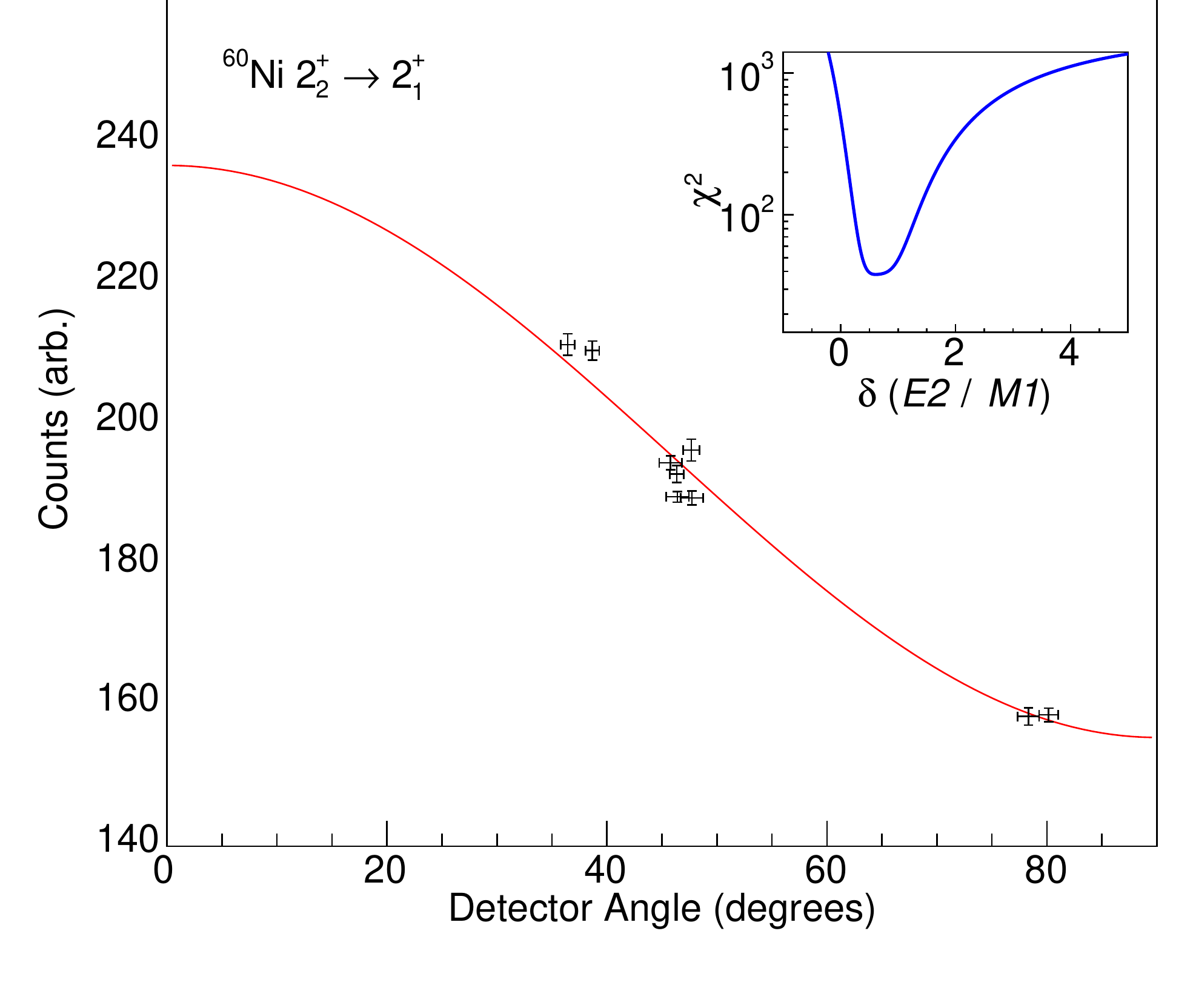}
\caption{Example $\gamma$-ray angular distribution for the \tra{2}{2}{2}{1} transition in $^{60}$Ni from the ($p,p'\gamma$) measurement. The inset shows the associated $\chi^2$ minimization curve.}
\label{fig:ang_dist}
\end{figure}
 
\begin{figure}[!ht]
	\centering
	\includegraphics[width=\linewidth]{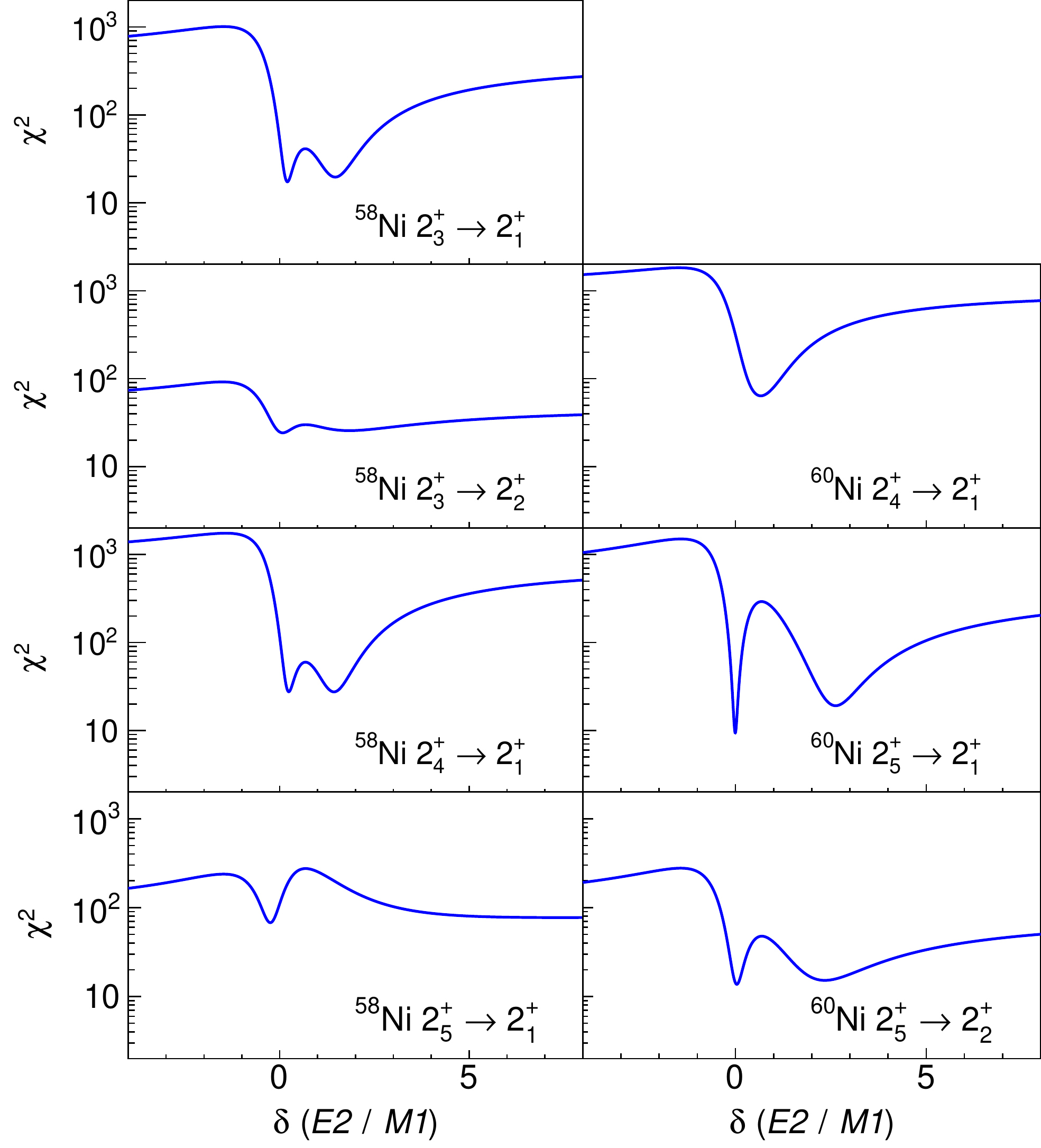}
	\caption{The $\chi^2$ plots for the sensitivity to the mixing ratio in the $\gamma$-ray angular distribution for \tra{2}{}{2}{} transitions observed in $^{58}$Ni (left) and $^{60}$Ni (right).}
	\label{fig:panel_chi}
\end{figure}

The results for $\delta (E2/M1)$ mixing ratios from this work are presented in Table \ref{tab:resultsDelta}. The values presented for the \tra{2}{2}{2}{1} transitions in the three isotopes have been discussed in our previous publication \cite{Evitts_PLB}.  The $\gamma$-ray angular distribution for the \tra{2}{2}{2}{1} transition in $^{60}$Ni from the ANU data is shown in Fig. \ref{fig:ang_dist}. The $\delta (E2/M1)$ mixing ratio of the 1321.2\,keV transition of $^{58}$Ni is from the UKAL data (Fig. \ref{fig:AD}), for the 826.06\,keV \tra{2}{2}{2}{1} transition in $^{60}$Ni the weighted mean of the values obtained in the ANU and UKAL measurements are used, and for the 1128.82\,keV \tra{2}{2}{2}{1} transition in $^{62}$Ni the weighted mean of our value from the ANU data and that reported in Ref. \cite{chakraborty2011} is used. The measurements for $\delta$($E2$/$M1$) mixing ratios of all other transitions reported here are from the ANU data.  The $\chi^2$ distributions for angular distribution data collected with CAESAR are shown in Fig. \ref{fig:panel_chi} with the corresponding results summarized in Table \ref{tab:resultsDelta}.  Two values are reported for some transitions as there are two minima in the $\chi^2$ plot, both of which are used in determining the $\rho^2 (E0)$, $B(M1$), and $B(E2$) values of the \tra{2}{}{2}{} transitions.  The majority of the new measurements, for which literature values are available, agree within 1$\sigma$ of the adopted values listed in the evaluated Nuclear Data Sheets \cite{NucData58, NucData60, NucData62}. There are also a number of new values from the present work, particularly in $^{60}$Ni.

The $\delta$ value of the 1791\,keV transition in $^{60}$Ni could not be measured due to intense background in the spectrum from the 1779\,keV $2^+ \rightarrow 0^+$ $\gamma$ ray of $^{28}$Si, which was observed in the CAESAR data only as a result of scattered protons striking the glass target chamber. The Super-e spectra do not display this contamination. The literature value for the mixing ratio was used in order to determine $\rho^2 (E0)$ of the 1791\,keV transition.

\begin{table}
\centering
\caption{Experimental $\delta$($E2$/$M1$) multipole mixing ratios determined in the present work. The columns $E_\gamma$ and $E_i$ are the transition and initial level energy, respectively. The $\delta$ values listed under NDS are taken from the evaluated Nuclear Data Sheets \cite{NucData58,NucData60,NucData62}.}
\label{tab:resultsDelta}
\begin{tabular}{llllll}
\hline
 &  &  &  & \multicolumn{2}{c}{$\delta(E2/M1)$} \\
 & Transition & E$_\gamma$ [keV] & E$_i$ [keV] & This work & NDS  \\ 
 \hline
$^{58}$Ni & \tra{2}{2}{2}{1} & 1321.2 & 2775.42 & -1.04$^{+0.07}_{-0.08}$ & -1.1(1)  \\
 & \tra{2}{3}{2}{1} & 1583.8 & 3037.86 & +0.20(4) & +0.21(3)   \\
 &  &  &  & +1.48(13) & +2.1$^{+1.6}_{-0.7}$   \\
 & \tra{2}{3}{2}{2} & 262.6 & 3037.86 & +0.07$^{+0.14}_{-0.10}$ & -0.03(5)  \\
 & \tra{2}{4}{2}{1} & 1809.5 & 3263.66 & +0.24(4) & +0.7(4)   \\
 &  &  &  & +1.42(10) &    \\
 & \tra{2}{5}{2}{1} & 2444.7 & 3898.8 & -0.11(4) & 0.0(1)   \\
 \hline
$^{60}$Ni & \tra{2}{2}{2}{1} & 826.06 & 2158.63 & +0.43(8) & +0.9(3)   \\
 & \tra{2}{3}{2}{1} & 1791.6 & 3123.69 &  & -0.21(4)   \\
 & \tra{2}{4}{2}{1} & 1936.9 & 3269.19 & +0.66(8) &     \\
 & \tra{2}{5}{2}{1} & 2060.58 & 3393.14 & -0.01(2) &     \\
 &  &  &  & +2.62$^{+0.16}_{-0.14}$ &   \\
 & \tra{2}{5}{2}{2} & 1234.51 & 3393.14 & +0.04(5) &     \\
 &  &  &  & +2.3$^{+0.4}_{-0.3}$ &     \\ 
 \hline
$^{62}$Ni & \tra{2}{2}{2}{1} & 1128.82 & 2301.84 & +3.1(1) & +3.19(11)   \\
 &  &  &  & -0.07(1) &    \\ 
 \hline
\end{tabular}
\end{table}

\subsection{B($M1$) and B($E2$) values}

From the new values of $\delta$($E2$/$M1$) obtained in this work, the reduced transition probabilities, $B(M1$) and $B(E2$), for each mixed transition were calculated as,

\begin{equation}
B(M1) = \left( \frac{1}{1+\delta^2} \right) \frac{3.17 \times 10^7}{E_\gamma^3 \cdot \tau_p \cdot (1 + \alpha_T)},
\end{equation}

\noindent and

\begin{equation}
B(E2) = \left(\frac{\delta^2}{1+\delta^2} \right) \frac{1.37 \times 10^{19}}{A^{4/3} \cdot E_\gamma^5 \cdot \tau_p  \cdot (1 + \alpha_T)},
\end{equation}

\noindent where $B(\lambda L$) is in Weisskopf units, $\tau_p$ is the partial mean lifetime in ps determined from the $\gamma$-ray branching ratio, $E_\gamma$ is the transition energy in keV, and $\alpha_T$ is the coefficient for all other possible decay modes including internal conversion and internal pair formation, typically taken from theory \cite{brookhaven1987procedures}.  The results are shown in Table \ref{tab:results} and compared to the adopted values in the Nuclear Data Sheets \cite{NucData58, NucData60, NucData62}, where available. The new measurements of mixing ratios allow a number of transition strengths to be determined for the first time.

\subsection{Internal conversion coefficients}

\begin{figure}[!ht]
	\centering
	\includegraphics[width=1.0\linewidth]{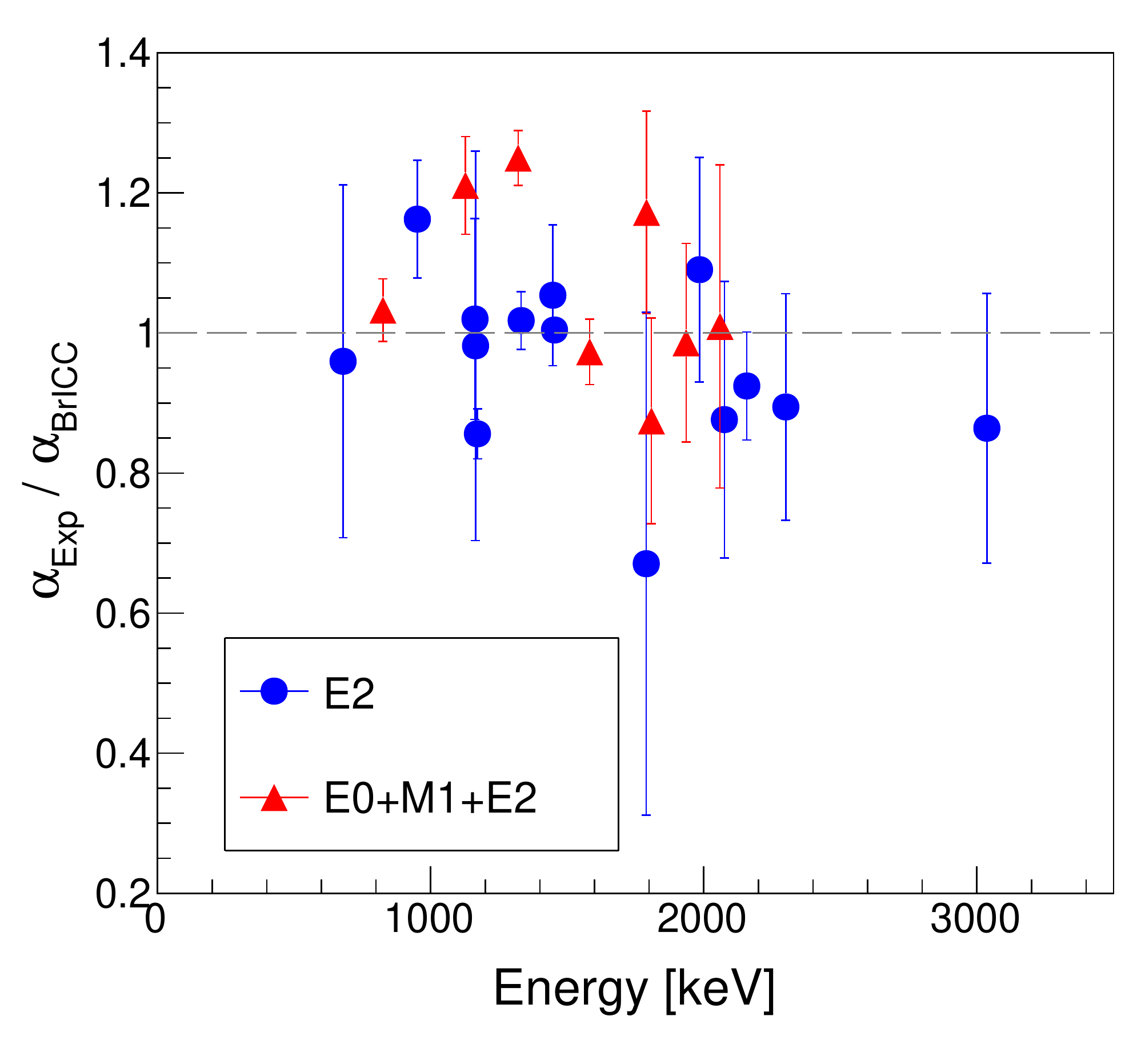}
	\caption{Ratio of experimental to theoretical K-shell internal conversion coefficients.  For $E0+M1+E2$ transitions the theoretical values are only for $M1+E2$ multipolarities and the experimental uncertainty in the $\delta$($E2$/$M1$) mixing ratio is included in the error bar.}
	\label{fig:all_alpha}
\end{figure}

The experimental K internal conversion coefficients (ICC) for $^{58, 60, 62}$Ni are listed in Table \ref{tab:results}. The uncertainties are dominated by the limited statistics of the electron spectra. The ratio of the experimental to theoretical ICC values for pure $E2$ and mixed ($E0$+$M1$+$E2$) multipolarity are shown in Fig. \ref{fig:all_alpha} as a function of transition energy. 
In the case of mixed ($E0$+$M1$+$E2$) transitions, the theoretical $\alpha_{BrICC}$ value used to construct the $\alpha_{Exp} / \alpha_{BrICC}$ ratio is calculated using the experimental $\delta(E2/M1)$ mixing ratio.
The experimental uncertainty in the mixing ratio is accounted for in the error bar. There is generally good agreement for the pure $E2$ transitions. 
Two transitions require further comment. The electron peak of the 952\,keV \tra{0}{2}{2}{1} transition in $^{60}$Ni overlaps with that of a 947\,keV transition reported in $^{60}$Cu, generated by the (p,n) reaction. Fitting of the $\gamma$-ray peak of the 1172\,keV, \tra{2}{1}{0}{1} transition in $^{62}$Ni is complicated by overlap with the 1164\,keV transition reported in the same nucleus. In these two cases, these contaminations in the experimental spectra prevented good agreement with the theoretical coefficient.
In a number of the mixed transitions, particularly the \tra{2}{2}{2}{1} transitions, there is significant $E0$ strength indicated by an $\alpha_{Exp} / \alpha_{BrICC}$ ratio greater than 1.

The \tra{0}{2}{0}{1} transitions in $^{60,62}$Ni, which have only been previously observed through internal pair formation measurements \cite{Passoja1981, Warburton197138}, are observed here by internal conversion decay. The ratio of the $E0$ conversion coefficients to the $E2$ conversion coefficient of the competing decay branch to the $2^+$ state, $q_k^2 = I^{E0}_k / I^{E2}_k$, derived from the previous work can be compared to the new data.  In $^{60}$Ni, the $q_k^2$ value was measured in the current work to be 0.079(8), which agrees well with the previously measured value of 0.074(16) \cite{Passoja1981}.  For $^{62}$Ni, the $q_k^2$ value was measured as 0.119(14), which only agrees with the previous value of 0.084(11) \cite{Passoja1981} within 2$\sigma$.

Comparison of measured $q^2$ values must consider the models used to evaluate the pair formation and $e^+e^-$ angular distributions, which can affect the calculated efficiency of a pair spectrometer through a dependence on the emission angles of the emitted particles.  
In the 1990s, models suitable for all elements were developed employing the distorted-wave Born approximation (DWBA) method, which includes relativistic effects, the spin orientation specified via magnetic substates, and the finite size of the nucleus \cite{Leinberger1998}.  
Earlier models had used the Born approximation with plane waves \cite{Warburton1964,Wilkinson1963,Warburton1963,Rose.76.1949,Hofmann1990}.
The theoretical $\alpha_\pi$ values and angular distributions of emitted particles differ considerably between the Born and DWBA approximations, particularly for magnetic transitions \cite{Leinberger1998}.
The previous measurements for Ni isotopes \cite{Passoja1981, Warburton197138} followed the formalism detailed in Refs. \cite{Warburton1964,Wilkinson1963,Warburton1963} for calculations of detection efficiency which could provide an explanation for agreement at only the 2$\sigma$ with the present  $^{62}$Ni result.

\subsection{$E0$ transition strengths}

\begin{figure*}[!ht]
	\centering
	\includegraphics[width=0.9\linewidth]{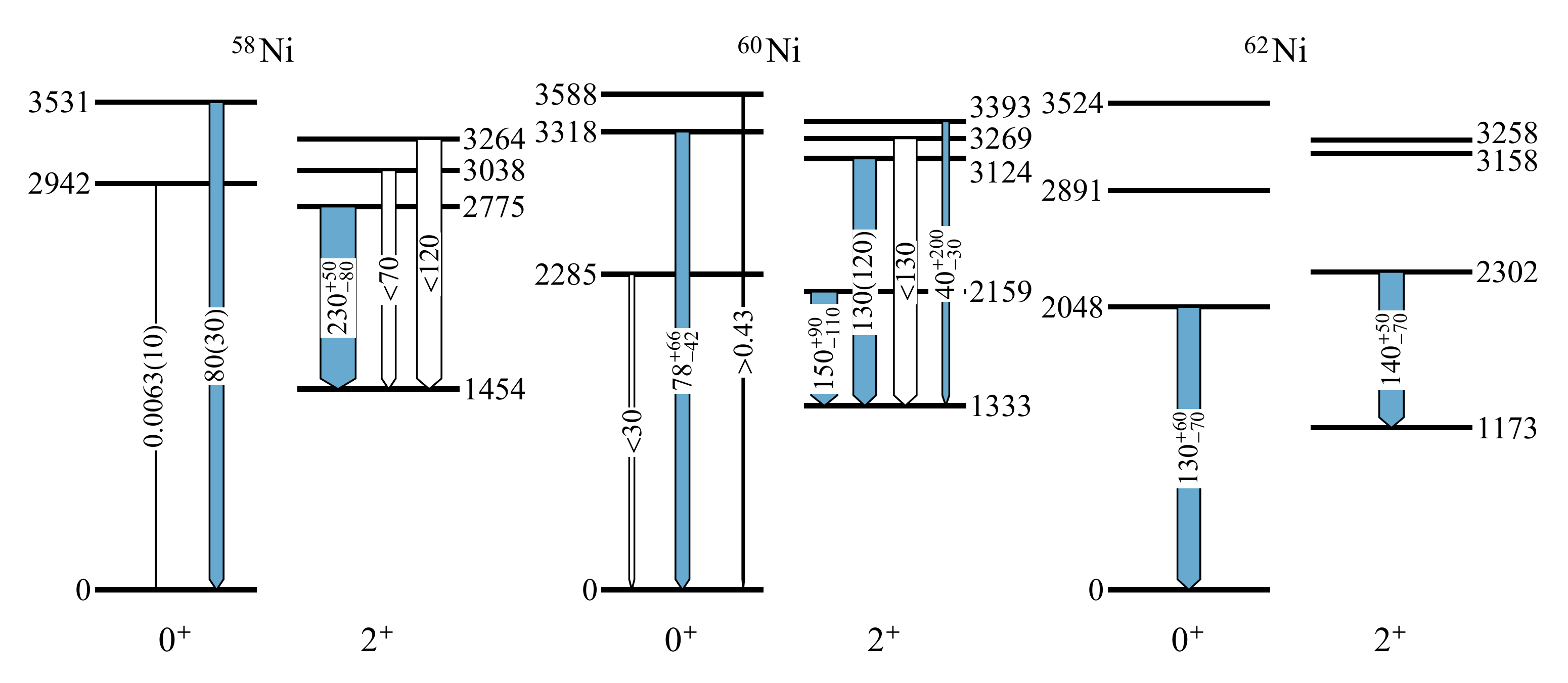}
	\caption{Experimental $\rho^2 (E0) \times 10^3$ values measured in this work, combined with previous literature values in $^{58, 60}$Ni \cite{Passoja1981, Warburton197138}. Unfilled transitions indicate that an upper limit has been determined. Level energies are shown in keV. The levels are grouped by their value of $J^\pi$ so that $E0$ transitions where $\Delta J=0$ appear vertically.}
	\label{fig:rho_summary}
\end{figure*}

Using the $\delta$($E2$/$M1$) mixing ratios (Section \ref{sect:MR}, Table \ref{tab:resultsDelta}) and internal conversion coefficients measured in this work, along with previously reported values from the literature \cite{NucData58, NucData60, NucData62}, the $E0$ transition strengths were determined and are shown in Table \ref{tab:results}.  Branching ratios were determined from the relative photon intensities reported in \cite{NucData58, NucData60, NucData62} in combination with the new values for mixing ratios and conversion coefficients.
For transitions where there are two solutions for the measured $\delta$($E2$/$M1$) mixing ratio, both values were used individually to obtain separate $\rho^2 (E0)$ values.  The results, along with the previously reported results, are summarized in Fig. \ref{fig:rho_summary}.  In $^{58}$Ni, many of the newly determined $E0$ transition strengths have upper limits.   In $^{60}$Ni, there is an upper limit on the 2285\,keV \tra{0}{2}{0}{1} transition strength because the half-life of the parent state has only a lower limit of 1.5\,ps \cite{NucData60}.  

The \tra{2}{}{2}{} $E0$ transition strengths found here are consistently large in all three of the Ni isotopes studied, particularly for the \tra{2}{2}{2}{1} transitions.  In almost all transitions, the dominant source of error is the small number of events observed in the e$^-$ spectra, particularly those from higher-lying states where only an upper limit could be obtained.

As has been previously discussed \cite{Heyde.83.1467,Wood1999}, large $E0$ strength is typically associated with differences in deformation and mixing between configurations. This condition appears to be the origin of the strong $E0$ transition between the third and first $0^+$ states in $^{58,60}$Ni \cite{Wood1999,Passoja1981}. These excited states are the ones observed to be strongly populated in 2-proton \cite{Evers1972} and alpha \cite{Fulbright1977} transfer reactions and, therefore, are interpreted as two-particle, two-hole (2p-2h) excitations across the $Z=28$ proton shell closure. In stark contrast, the $E0$ transition strength between the $0^+_2$ state (very weakly populated in transfer) and the ground state is observed to be very weak \cite{Warburton197138}.
These 2p-2h `intruder' configurations are usually associated with deformation and collectivity with the quadrupole neutron-proton interaction being a key driver in the development of such behavior. This creates a shape coexistence scenario with strong $E0$ transitions between the deformed 2p-2h intruder states and the spherical states.
From the pattern of $E0$ transition strength, it appears that the 2p-2h state is the $0^+_2$ state in $^{62}$Ni but transfer data are not available to support this assignment. In light of this shape coexistence interpretation for the pattern of $E0$ strength between the $0^+$ states, the strong $E0$ transitions observed between the lowest-lying $2^+$ states are even more surprising. The $2^+_2$ levels lie well below the excited $0^+$ states and, therefore, exclude the possibility that these excitations are built on the 2p-2h configuration.

The microscopic model of Brown {\it et al.} \cite{Brown.95.011301} does not reproduce the new experimental results for \tra{2}{}{2}{} transitions, although this model is successful in reproducing $E0$ transition strengths in \tra{0}{}{0}{} cases. In $^{58}$Ni, the calculated $\rho^2$(E0) value for the 0$^{+}_{2} \rightarrow 0^{+}_{1}$ transition was much larger than the experimental value.
A significant improvement in agreement was achieved through a remixing of the 0$^{+}_{2}$ - 0$^{+}_{3}$ and 2$^{+}_{2}$ - 2$^{+}_{3}$ states. The calculated $\rho^2$(E0) for the remixed 0$^{+}$ states was about a factor of two smaller than in experiment (comparable to the level of agreement achieved in the other nuclei studied in Ref. \cite{Brown.95.011301}). This observation highlights the sensitivity of $E0$ transition strengths to configuration mixing and to small components of the wavefunctions for the states involved in the transition.
The $B(M1$) and $B(E2$) values, including the ones newly obtained in the present work, as well as moments, are also well reproduced in this shell-model framework. The largest \tra{2}{2}{2}{1} $E0$ transition strength calculated using the microscopic model is 6~milliunits in $^{58}$Ni, while the transitions between higher-lying $2^+$ states are predicted to be even weaker. Further details can be found in Ref. \cite{Evitts_PLB}. Certainly, large-basis shell-model calculations would be illuminating whether or not they succeed in describing the observed $E0$ strength.

\begin{figure}[!ht]
	\centering
	\includegraphics[width=0.9\linewidth]{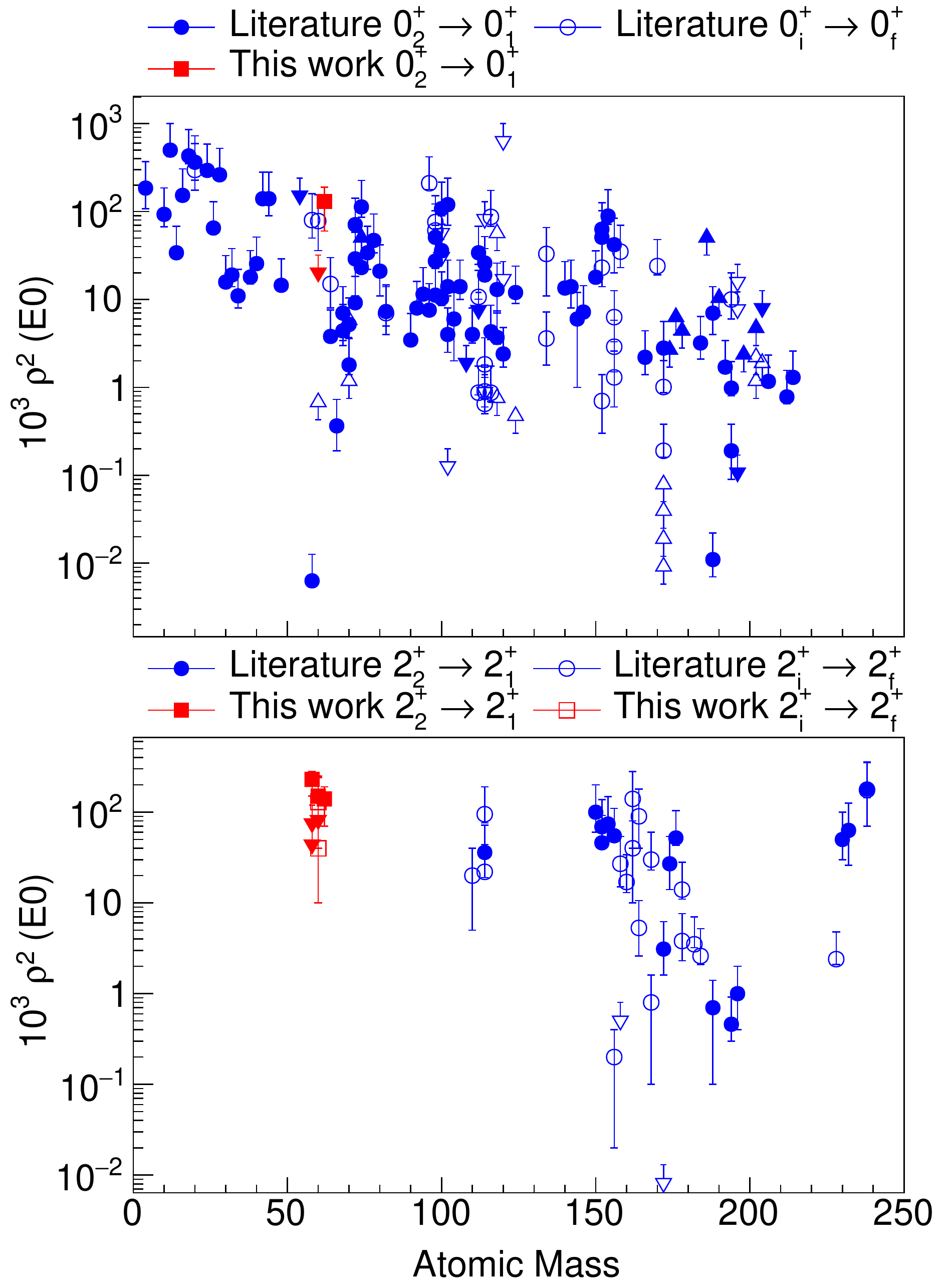}
	\caption{The known $\rho^2 (E0)$ values for (a) \tra{0}{i}{0}{f} and (b) \tra{2}{i}{2}{f} transitions as a function of atomic mass.  Upper/lower limits are shown as triangles with the error bar indicating the relevant limit. The data are from the most recent compilations by Kib\'{e}di \cite{Kibedi2005} and Wood \cite{Wood1999}.}
	\label{fig:all_rho}
\end{figure}

The values obtained in this work are compared to other $E0$ transition strengths across the chart of nuclides in Fig. \ref{fig:all_rho}, where the filled data points are for \tra{0}{2}{0}{1} and \tra{2}{2}{2}{1} transitions, while the open data points are other \tra{0}{i}{0}{f} and \tra{2}{i}{2}{f} transitions.  It can be clearly seen that the \tra{2}{}{2}{} $E0$ transitions in these stable Ni isotopes have considerable strength and are among the largest measured. 
Based on a shell-model approach one can apply a `single-particle' scaling factor of $A^{2/3}$ to $E0$ strength, which should provide values that are independent of mass \cite{Voinova1986,Wood1999}. When this is done, the observed Ni values remain amongst the largest, along with the \tra{2}{2}{2}{1} transition in $^{238}$Pu. 
In the case of \tra{0}{}{0}{} transitions this scaling was suggested to perhaps be insufficient \cite{Kibedi2005} as a downward trend in $E0$ transition strength was still present as a function of mass number. The low number of experimental values available for \tra{2}{}{2}{} $E0$ transitions prevents global conclusions on systematic behavior from being drawn at this time.

On the experimental side, it would be of value to measure $E0$ transition strengths for other \tra{2}{}{2}{} transitions in order to build a comprehensive picture of the behavior of $E0$ transition strengths in atomic nuclei. This enterprise will require precise measurements of lifetimes, branching ratios, and mixing ratios along with conversion coefficients: such measurements are challenging, but feasible, and will illuminate an important aspect of nuclear structure that is poorly characterized at present.

\begin{sidewaystable*}
\centering
     \rule[-.3\baselineskip]{0pt}{0.3\textheight}
\caption{The experimental multipole $\delta$($E2$/$M1$) mixing ratios (see Table \ref{tab:resultsDelta}), K internal conversion coefficients ($\alpha_K$), electric monopole transition strengths $\rho^2 (E0)$, and $E0$ matrix elements $M(E0)$ obtained in this work. $q_k^2$ is the ratio of the $E0$ conversion coefficient to the $E2$ conversion coefficient of the competing decay branch. Transition strengths of $B(M1$) and $B(E2$) are also given as determined from the present work. Comparisons are made to the adopted values in the Nuclear Data Sheets where available \cite{NucData58, NucData60, NucData62}.  The columns $E_{trans}$ and $E_i$ are the transition and parent level energy respectively, and $T_{1/2}$ is the half-life of the parent state.}
\label{tab:results}
\begin{tabular}{cccccccccccccc}
\hline
 &  &  &  &  &  &  &  &  &  & \multicolumn{2}{c}{This work} & \multicolumn{2}{c}{Literature} \\
 & Transition & $E_{trans}$ & $E_i$ & $T_{1/2}$ & $\delta(E2/M1)$ & $\alpha_K \times 10^4$  & $q^2_K$ & $\rho^2(E0)$ & $M (E0$) & $B(M1$) & $B(E2$) & $B(M1$) & $B(E2$) \\
 & & ($keV$) & ($keV$) & ($ps$) & & & & ($\times 10^3$) & ($fm^2$) & ($W.u.$) & ($W.u.$) & ($W.u.$) & ($W.u.$) \\
 \hline
$^{58}$Ni & \tra{0}{2}{0}{1} & 2942.6 & 2942.6 & 1460(140) & - & - & 0.65(10) & 0.0063(10) &  &  &  &  & \\
 & \tra{0}{3}{0}{1} & 3531.1 & 3531.1 & 0.19(6) & - & - & 0.27(4) & 80(30) &  &  &  &  & \\
 & \tra{2}{2}{2}{1} & 1321.2 & 2775.42 & 0.60$^{+0.19}_{-0.12}$ & -1.04$^{+0.07}_{-0.08}$ & 1.38(3) & 0.46(7) & 230$^{+50}_{-80}$ & 10.3$^{+1.1}_{-2.0}$ & 7.3$^{+1.6}_{-2.4} \times 10^{-3}$ & 9$^{+2}_{-3}$ & 0.011$^{+0.003}_{-0.004}$ & 15$^{+4}_{-5}$ \\
 & \tra{2}{3}{2}{1} & 1583.8 & 3037.86 & 0.057(8) & +0.20(4) & 0.72(3)  & \textless 0.7 & \textless 70 & \textless 6 & 0.055(8)  & 1.7(7)  & 0.055(8) & 1.8(6) \\
 &  &  &  &  & +1.48(13) & & \textless 0.01 & \textless 22 & \textless 3 & 0.018(4) & 30(5) &  &  \\
 & \tra{2}{3}{2}{2} & 262.6 & 3037.86 & 0.057(8) & +0.007$^{+0.014}_{-0.010}$ &  &   &  &  & 0.21(5) & 0.3$^{+1.2}_{-0.3}$ & 0.21(5) & 5$^{+18}_{-5}$ \\
 & \tra{2}{4}{2}{1} & 1809.5 & 3263.66 & 0.037(5) & +0.24(4) & 0.52(9)  & \textless 1.3 & \textless 120 &  \textless 7 & 0.037(6) & 1.3(5) & 0.027(11) & 8(6) \\
 &  &  &  &  & +1.42(10) & & \textless 0.05 & \textless 80 & \textless 5 & 0.013(2) & 15(2) &  &  \\
 & \tra{2}{5}{2}{1} & 2444.7 & 3898.8 & --- & -0.11(4) & & & & & 0.049(13) & 0.19(5) & 0.050(13) &  \\ 
 \hline
$^{60}$Ni & \tra{0}{2}{0}{1} & 2284.8 & 2284.8 & \textgreater 1.5 & - & - &  0.079(8) & \textless 30  & \textless 4 \\
 & \tra{0}{3}{0}{1} & 3317.8 & 3317.8 & 0.24$^{+0.28}_{-0.11}$ & - & - & 0.29(3) & 78$^{+66}_{-42}$ &  &  &  &  & \\
 & \tra{0}{4}{0}{1} & 3587.7 & 3587.7 & \textless 40 & - & - & 1.26(20) & \textgreater 0.43 &  &  &  &  & \\
 & \tra{2}{2}{2}{1} & 826.06 & 2158.63 & 1.28$^{+0.74}_{-0.35}$ & +0.43(8) & 3.0(1)  & 0.4$^{+0.2}_{-0.3}$ & 150$^{+90}_{-110}$ & 9$^{+2}_{-4}$ & 0.022$^{+0.007}_{-0.013}$ & 11$^{+5}_{-8}$ & 0.031(13) & 70(40)\\
 & \tra{2}{3}{2}{1} & 1791.6 & 3123.69 & 0.23$^{+0.17}_{-0.10}$ & -0.21(4) & 0.69(9) & 4(3)  & 130(120) & 8$^{+3}_{-6}$ &  &  & 0.013$^{+0.006}_{-0.010}$ & 0.34$^{+0.20}_{-0.28}$  \\
 & \tra{2}{4}{2}{1} & 1936.9 & 3269.19 & 0.071(21) & +0.66(8) & 0.52(7) & \textless 0.4 & \textless 130 & \textless 8 & 0.013(4) & 2.8(10) &  &  \\
 & \tra{2}{5}{2}{1} & 2060.58 & 3393.14 & 0.13$^{+0.06}_{-0.04}$ & -0.01(2) & 0.48(11)  & 150$^{+1000}_{-150}$ & 40$^{+200}_{-30}$ & 4$^{+6}_{-2}$ & 0.016$^{+0.005}_{-0.008}$ & 0.001$^{+0.003}_{-0.001}$ &  &  \\
 &  &  &  &  & +2.62$^{+0.16}_{-0.14}$ & & \textless 0.3 & \textless 200 & \textless 8 & 2.1$^{+0.7}_{-1.0} \times 10^{-3}$ & 6$^{+2}_{-3}$ &  &  \\
 & \tra{2}{5}{2}{2} & 1234.51 & 3393.14 & 0.13$^{+0.06}_{-0.04}$ & +0.04(5) &   &  &  & & 0.009$^{+0.003}_{-0.005}$ & 0.02$^{+0.05}_{-0.02}$ &  &  \\
 &  &  &  &  & +2.3$^{+0.4}_{-0.3}$ &  &   &  & & 1.5$^{+0.7}_{-0.9}$$\times$$10^{-3}$ & 10$^{+3}_{-5}$ &  &  \\ 
 \hline
$^{62}$Ni & \tra{0}{2}{0}{1} & 2048.68 & 2048.68 & 0.92$^{+0.67}_{-0.23}$\footnote{A weighted average is taken from Refs \cite{NucData62} and \cite{chakraborty2011}.} & - & - &  0.084(11) & 130$^{+60}_{-70}$  & 8.1$^{+1.7}_{-2.6}$ \\
 & \tra{2}{2}{2}{1} & 1128.82 & 2301.84 & 0.67$^{+0.20}_{-0.14}$ & +3.1(1) & 1.95(11)  & 0.22(7) & 140$^{+50}_{-70}$ & 8.4$^{+1.4}_{-2.5}$ & 9$^{+2}_{-3} \times 10^{-4}$ & 13$^{+3}_{-4}$ & 1.06$^{+0.18}_{-0.3} \times 10^{-3}$ & 14.9$^{+2.4}_{-4.2}$\\
 \hline
\end{tabular}
\end{sidewaystable*} 

\section{Conclusion \label{sect:conclusion}}

In this work, the $E0$ transition strengths between $J^\pi=2^+$ states were measured for three of the stable Ni isotopes, $^{58, 60, 62}$Ni.  These new values were obtained through measurements of the $\delta$($E2$/$M1$) mixing ratio and internal conversion coefficients combined with level lifetimes. The new data also allow a number of $B(M1$) and $B(E2$) values to be determined for the first time. The $E0$ transition strengths between $0^+$ states were measured using internal conversion electron spectroscopy for the first time and compare well to previous results from internal pair formation spectroscopy \cite{Warburton197138,Passoja1981}.

As was discussed in our previous publication \cite{Evitts_PLB}, this work contains the first reported $E0$ transition strength information for \tra{2}{}{2}{} transitions in nuclei with $A<100$. These also represent the first evaluation of \tra{2}{}{2}{} $E0$ strengths in nuclei with spherical ground states, as previous research focused on the lanthanide region. The explanation of the significant $E0$ strength observed in these isotopes should be the focus of future theoretical efforts.

\begin{acknowledgments}
\sloppy
We would like to thank the technical staff of the Heavy Ion Accelerator Facility at the Australian National University, and in particular Justin Heighway for preparing the nickel targets. We thank B.A.~Brown and S.R.~Stroberg for useful discussions related to this work. A.B.G. is grateful for support from the Department of Nuclear Physics of the Australian National University. Support for the ANU Heavy Ion Accelerator Facility operations through the Australian National Collaborative Research Infrastructure Strategy (NCRIS) program is acknowledged.
This work was supported in part by the Natural Sciences and Engineering Research Council of Canada (NSERC); the U.S. National Science Foundation, Grant No. PHY-1606890; and by the Australian Research Council Discovery Grants DP140102986 and FT100100991. TRIUMF receives funding via a contribution agreement through the National Research Council Canada.
\end{acknowledgments}

\bibliography{main}

\end{document}